\documentclass[fleqn,usenatbib]{mnras}

\usepackage{newtxtext,newtxmath}

\usepackage[T1]{fontenc}
\usepackage{multicol}

\DeclareRobustCommand{\VAN}[3]{#2}
\let\VANthebibliography\thebibliography
\def\thebibliography{\DeclareRobustCommand{\VAN}[3]{##3}\VANthebibliography}

\usepackage{graphicx}	
\usepackage{amsmath}	
\usepackage{soul}

\newcommand{\src}{HERS1}

\newcommand{\HST}{\textit{HST}}

\newcommand{\unitjybeam}{$\rm{Jy\,\,{\rm beam}^{-1}}$}

\newcommand{\datapol}{{$\rm D_{2}$}}
\newcommand{\datares}{{$\rm D_{1}$}}

\newcommand{\pronto}{\textsc{pronto}}

\newcommand{\multinest}{\textsc{MultiNest}}

\newcommand{\data}{\bmath{d}}
\newcommand{\source}{\bmath{s}}
\newcommand{\sourcemap}{\bmath{s}_{\rm MP}}
\newcommand{\sourcemapT}{\bmath{s}^T_{\rm MP}}
\newcommand{\lambdas}{\bmath{\lambda}_{\source{}}}

\newcommand{\lensop}{\bmath{\mathsf{L}}}
\newcommand{\dirty}{\bmath{\mathsf{D}}}
\newcommand{\lensparam}{\bmath{\eta}}
\newcommand{\noise}{\bmath{n}}
\newcommand{\invcov}{\bmath{\mathsf{C}}^{-1}}
\newcommand{\faraday}{\bmath{\mathsf{S}}}
\newcommand{\regul}{\bmath{\mathsf{R}}}

\newcommand{\posang}{\theta_{q}}

\title[An ordered magnetic field at redshift 2.6]{A grand-design spiral galaxy with an ordered magnetic field at redshift 2.6 as resolved with ALMA and gravitational lensing}

\author[W. de Roo et al.]{
W. de Roo,$^{1}$\thanks{E-mail: roo@mpa-garching.mpg.de}
S. Vegetti,$^{1}$
D.~M. Powell,$^{1}$
S.~W. Ndiritu$^{1,2}$,
R. Pakmor$^{1}$
and J.~P. McKean$^{2,3,4}$
\\
$^{1}$Max Planck Institute for Astrophysics, Karl-Schwarzschild Straße 1, D-85748, Garching bei München, Germany\\
$^{2}$Kapteyn Astronomical Institute, University of Groningen, P.O. Box 800, 9700 AV Groningen, The Netherlands\\
$^{3}$South African Radio Astronomy Observatory (SARAO), P.O. Box 443, Krugersdorp 1740, South Africa\\
$^{4}$Department of Physics, University of Pretoria, Lynnwood Road, Hatfield, Pretoria, 0083, South Africa\\
}

\date{Accepted XXX. Received YYY; in original form 2024 December 11}

\pubyear{\the\year{}}

\begin{document}
\label{firstpage}
\pagerange{\pageref{firstpage}--\pageref{lastpage}}
\maketitle

\begin{abstract}
Magnetic fields play an important role in the evolution of galaxies and in shaping the dynamics of their inter-stellar medium. However, the formation history of magnetic fields from initial seed-fields to well-ordered systems is not clear. Favoured scenarios include a turbulent dynamo that amplifies the field, and a mean-field dynamo that organizes it. Such a model can be tested through observing the magnetic-field structure of galaxies in the early Universe given the relative formation time-scales involved. Here, we combine the high angular resolution of the Atacama Large Milli-metre Array (ALMA) and gravitational lensing to resolve the magnetic field structure of a 4-kpc in extent grand-design spiral when the Universe was just 2.6 Gyr old. We find that the spiral arm structure, as traced by the heated dust emission, is coincident with the linearly polarized emission, which is consistent with a highly ordered magnetic field. The time-scale needed to produce such an ordered field is likely within at least several rotations of the disk. Our study highlights the importance of combining the long baselines of ALMA and gravitational lensing to resolve the structure of galaxies at cosmologically interesting epochs.
\end{abstract}


\section{Introduction}

Magnetic fields are believed to play a crucial role in the evolution of galaxies, shaping the dynamics of the turbulent interstellar medium (ISM), regulating star formation and governing the distribution of cosmic rays within the ISM \citep{beck_2009}. Therefore, understanding the spatial structure of magnetic fields is needed to test various models for how they form and co-evolve within galaxy-scale environments, and determine how they impact galaxy formation and evolution. Observing the structure of magnetic fields in galaxies is commonly done indirectly through measurements of the polarized thermal emission from magnetically aligned dust grains \citep[e.g.][]{andersson_2015}, or via Faraday rotation measurements of polarized synchrotron emission \citep[e.g.][]{beck_2019}. Throughout this work we adopt the naming convention for the different magnetic field components from \citet{beck_2013}.

In the Local Universe, where the magnetic field structure can be well resolved, recent polarimetric far-infrared (FIR) observations of thermal dust emission in fourteen nearby galaxies suggest that kpc-scale ordered magnetic fields are common \citep{lopez-rodriguez_2022, lopez-rodriguez_2024}. This result is in good agreement with studies of the low-frequency synchrotron emission for a similar sample of galaxies \citep{mulcahy_2014,heesen_2024}. At higher redshifts, such studies have been challenging due to the faint emission levels and the poorer angular resolution. Whereas the Square Kilometre Array (SKA) and the upgraded Low Frequency Array (LOFAR2.0) are expected to significantly improve our capability of studying magnetic fields in distant galaxies, pushing beyond redshift two will remain challenging.

Strong gravitational lensing represents a unique opportunity to investigate the role of magnetic fields in the distant Universe. \citet{mao_2017} derived the first measurement of a $\mu G$ coherent magnetic field in a gravitational lensing galaxy (a spiral at redshift 0.439). Recently, \citet{ndiritu_2024} have proposed a new lens modelling approach that opens up the opportunity to extend this type of measurement to many more gravitational lens systems. By combining the sensitivity and resolution of the Atacama Large Milli-metre Array (ALMA) and strong gravitational lensing, the first resolved studies of the ordered magnetic fields in dusty star-forming galaxies (DSFGs) have been carried out \citep{geach_2023,chen_2024}. One of the most promising systems for such studies is HERS J020941.1+001557 (referred to as 9io9 in \citet{geach_2023}, but \src{} from hereon), which features a DSFG at redshift $z_{\rm src}=2.553$ that is gravitationally lensed by a foreground elliptical galaxy at redshift $z_{\rm lens}=0.2$ \citep{geach_2015}. Previously, \citet{geach_2023} analyzed full-polarization ALMA observations of \src{}, finding a 5 kpc-scale ordered magnetic field that is oriented parallel to the gaseous disk as the most likely configuration, under the assumption of a constant magnetic field in the source plane. However, that analysis was limited by the 1-arcsec beam size of the observations, which provides only several resolution elements across the lensed emission, and a simple model to describe the source surface brightness and magnetic field structure.

Here, we build on that previous study by presenting an analysis of higher-resolution full-polarization ALMA observations of the same system. We apply a novel lens modelling method to the data, which is done directly in the visibility space, and reconstructs the source-plane Stokes IQUV emission simultaneously. The pixellated model for the source surface brightness distribution allows us to reconstruct its properties with minimal prior assumptions on its morphology and/or polarization state. These polarization data are also combined with new high-resolution imaging of the Stokes I emission, which gives a detailed reconstruction of the dust continuum morphology in the source plane;  our analysis provides the most detailed view yet of the magnetic-field structure within a galaxy outside of the Local Universe.

Throughout, we assume the \citet{planck_collaboration_2016} cosmology.

\section{ALMA observations}\label{sec:obs}

    \begin{table*}
    	\centering
    	\caption{An overview of the observational parameters of the two datasets, where $\theta_{\rm maj}$, $\theta_{\rm min}$ and $\theta_{\rm PA}$ indicate the major axis, minor axis and position angle of the restoring beam, respectively. The position angle is measured east of north. The rms map noise of the residual visibilities is $\sigma$.}
    	\label{tab:obs}
    	\begin{tabular}{llllllllll} 
    		\hline
    		Project code & Shorthand & Band & Freq. range (GHz) & $\theta_{\rm maj}$ (mas) & $\theta_{\rm min}$ (mas) & $\theta_{\rm PA}$ ($\deg$) & $\sigma$ ($\mu$\unitjybeam{}) & Stokes  \\
    		\hline
                2023.1.01585.S & \datares{} & 7 & 341.63 - 357.78 & $53$ & $50$ & $-62.5$ & 24.9 & I  \\
    		2021.1.00458.S & \datapol{} & 7 & 335.52 - 351.45 & $460$ & $360$ & $-56.8$ & 26.0 & I, Q, U, V  \\
    		\hline
        \end{tabular}
        \label{tab:alma}
    \end{table*}

   We have obtained two publicly available ALMA datasets for our analysis of \src{}, which are summarized in Table~\ref{tab:alma}. These data were reduced using standard procedures, unless otherwise stated, within the Common Astronomy Software Applications (CASA; \citealt{CASA_2022}).

   The first ALMA dataset (2023.1.01585.S; PI: McKean; \datares{} from hereon) was observed on 2023 October 2 and 4 with a total on-source integration time of 47 min using the 12-m array in the C-8 configuration (8.5 km maximum baseline). The data were taken through the XX and YY linear correlations, and so, only provide a Stokes I image of the target. Four spectral windows centred on 342.6, 344.6, 354.6 and 356.6 GHz, each with 2 GHz bandwidth, were used. This results in a central observing frequency of 349.6 GHz, which corresponds to a rest-frame wavelength of 241.5~$\mu$m. The rms map noise of the naturally weighted residual visibilities is 24.9 $\mu$\unitjybeam{}, with a synthesized restoring beam of $53\times50$~mas at a position angle of $-62.5$~deg. The second ALMA dataset (2021.1.00458.S; PI: Chen; \datapol{} hereafter) was observed on 2022 September 5 and 7 in full polarization mode, allowing visibilities in Stokes I, Q, U and V to be formed. The total on-source integration time was 181 min. Similar to above, the data were taken in four spectral windows, although at different central frequencies of 336.5, 338.5, 348.5 and 350.5 GHz; the 6 GHz offset in the frequencies of the two datasets will have no impact on our analysis. The rms map noise of the naturally weighted residual visibilities is 26.0 $\mu$\unitjybeam{}, with a synthesized restoring beam of $0.46\times0.36$~arcsec at a position angle of $-56.8$~deg.

   The ALMA pipeline data products for the calibrators and the target were inspected for quality and any residual radio frequency interference was flagged. We then carried out phase-only self-calibration by iteratively creating a model for the sky surface brightness distribution until no improvement was seen in the structure of the emission, or the noise properties of the residual visibilities. The noise and data weights for \datares{} and \datapol{} were recomputed prior to lens modeling.

\begin{figure}
    	\includegraphics[width=1\columnwidth]{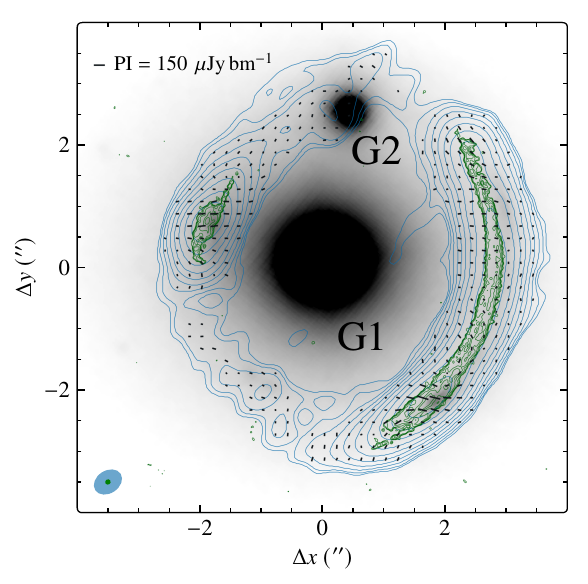}
        \caption{\HST{}/WFC3 F125W image of \src{}. The green contours show the naturally-weighted emission (de-convolved) from ALMA band 7 observations at 50 mas angular resolution (data set \datares{}). The light blue contours show the naturally weighted Stokes I image from data set \datapol{}. Contours for both observations start at $3\sigma$ and increase in steps of $1\sigma, 2\sigma, 4\sigma \ldots$. The $1\sigma$ level for \datares{} is at 25 $\mu$\unitjybeam{} and at 26 $\mu$\unitjybeam{} for \datapol{}. The light blue ellipse in the bottom left corner indicates the shape and orientation of the restoring beam for \datapol{}, the green ellipse inside indicates the restoring beam size for \datares{}. The black quivers show the polarization angles measured from Stokes Q and U of \datapol{} at positions where the corresponding Stokes I emission is above the 5-$\sigma$ level. The polarized intensity is proportional to the quiver length. The quiver in the top left corner corresponds to a polarized intensity of 150 $\mu$\unitjybeam{}.}
        \label{fig:hubblealma}
    \end{figure}

    Fig.~\ref{fig:hubblealma} shows the naturally weighted contours of \datares{} in green, overlayed on archival \HST{}/WFC3 data in the F125W filter (GO-15475; PI: H. Nayyeri). The emission at mm wavelengths appears slightly offset from the arc and counter-image in the F125W data. The counter image is resolved and reveals thermal dust emission in elongated regions of the background galaxy that extend from the main body, resembling spiral arms. The highly magnified arc, which is several arcsec in length, stretches part of the source along the Einstein ring. The centre of the background galaxy appears towards the bottom part of the main arc. The naturally weighted and deconvolved Stokes I image from \datapol{} is also shown as light blue contours in Fig.~\ref{fig:hubblealma}. The contours show that on these sub-arcsec scales the lensed emission is stretched to form a full Einstein ring at mm wavelengths, similar to what can be seen in the data from the \HST{}. The main difference between the low- and high-angular-resolution datasets is a 10-$\sigma$ peak in the \datapol{} contours to the south that appears to correspond to a slightly brighter region in the \HST{} data at that position. This may correspond to a second background source at the same redshift, as suggested by \citet{liu_2022}, or a low surface brightness extension to the main source. To the north of the Einstein ring, the companion lensing galaxy, denoted as G2, is located exactly on top of the ring. The brightness contours show an excess of mm emission at the 10-$\sigma$ level, just east of G2, which is due to localized splitting of the Einstein ring into multiple images at that location.

\section{Method}

To infer the lens mass distribution and the (polarized) surface brightness distribution of the background source, we make use of the lens modelling code \pronto{} \citep[e.g.][]{vegetti_2009, rybak_2015, rizzo_2018, ritondale_2019, powell_2021, ndiritu_2024}. Although originally developed for grid-based modeling of optical data, \pronto{} now implements a Bayesian approach for visibility-space modelling of polarized interferometric data \citep{powell_2021, ndiritu_2024}, where the data takes the form of a complex vector $\data{}$.

\subsection{Data model}

The ALMA receivers have linear feeds, which measure the orthogonal polarization states of incident electromagnetic radiation $e_{\rm X}$ and $e_{\rm Y}$. These measurements can be combined to form four complex correlation products that constitute the coherency vector
\begin{equation}
    \bmath{d}_{\rm XY} = \left(
    \langle e^{}_{\rm X}e^*_{\rm X}\rangle,
    \langle e^{}_{\rm X}e^*_{\rm Y}\rangle,
    \langle e^{}_{\rm Y}e^*_{\rm X}\rangle,
    \langle e^{}_{\rm Y}e^*_{\rm Y}\rangle
    \right)^T.
\end{equation}
The data $\data{}$ is obtained after the coordinate transformation $\data{}=\bmath{\mathsf{T}}\data{}_{\rm XY}$, where $\bmath{\mathsf{T}}$ is a complex-valued matrix that transforms the coherency vector from a linear representation to the abstract Stokes basis \citep{hamaker_1996}. The data $\data{}$ thus consists of $N_{\rm vis}\times N_{\rm corr}$ complex numbers, where $N_{\rm vis}$ is the total number of visibilities and $N_{\rm corr}$ the number of polarization products.

In similar fashion we can write down an expression for the measurement of the complex surface brightness of a polarized background source, $\source{}_{\rm XY}$. We relate the coherency vector $\source{}_{\rm XY}$ to the real-valued surface brightness distribution in Stokes coordinates $\source{}$ by the transformation
\begin{equation}
    \source{} =
    \left(
    \bmath{I}, \bmath{Q}, \bmath{U}, \bmath{V}
    \right)^T
    = \bmath{\mathsf{F}}_{\rm XY}\,\source{}_{\rm XY},
\end{equation}
where $\bmath{\mathsf{F}}_{\rm XY}$ is the feed operator \citep{ndiritu_2024}. We emphasize that $\source{}$, in contrast to $\data{}$, is not a complex quantity, but a real-valued description of the source surface brightness in the Stokes representation. Furthermore, we note that our notation deviates from \citet{ndiritu_2024} since we integrate over all the channels with multi-frequency synthesis, therefore not requiring channel indices.

The effects of gravitational lensing are captured in the lensing operator $\lensop{}(\lensparam{})$, which relates pixel coordinates on the image plane to an adaptive Delaunay-tessellated grid in the source plane, on which we represent the source surface brightness $\source{}$ \citep{vegetti_2009}. The lensing operator is a function of the parameters that describe the lens mass distribution, which we write as the vector $\lensparam{}$. The model for the data then takes the form
\begin{equation}\label{eq:model}
    \data{}  = \dirty{}\faraday{}\lensop{(\lensparam{})}\source{} + \noise{},
\end{equation}
where the NUFFT operator $\dirty{}$ accounts for the instrument response of the interferometer, the Faraday screen operator $\faraday{}$ includes the effects of Faraday rotation along the line of sight, and $\noise{}$ is assumed to be uncorrelated Gaussian noise with covariance $\invcov{}$ \citep{powell_2021, powell_2022}.

The Faraday screen operator was introduced to \pronto{} by \citet{ndiritu_2024} and accounts for the differential rotation of the polarization angle due to the interstellar medium of the lensing galaxy. Due to the high frequency of the ALMA observations in \datapol{} we assume negligible Faraday rotation and depolarization effects, which are proportional to $\lambda^2$. For typical rotation measure values in galaxies on the order of 100 rad m$^{-2}$ \citep{beck_2013} the observed Faraday rotation at 350 GHz would be less than $10^{-2}\deg$, so all measured polarized signal will be considered intrinsic to the background source. Therefore, we omit the Faraday screen operator from further equations.

\subsection{Bayesian inference}
\begin{figure*}
    \includegraphics[width=1\linewidth]{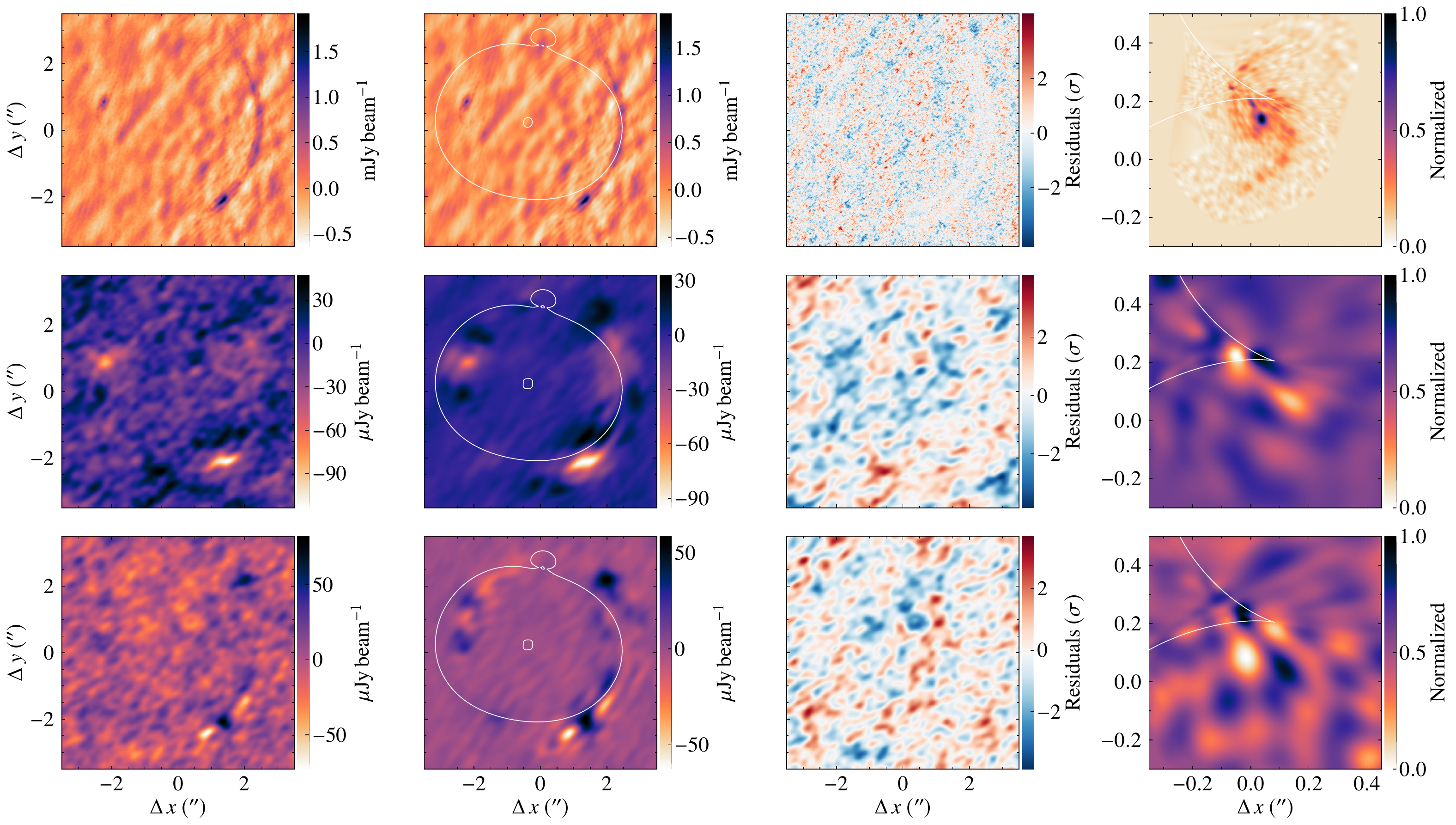}
    \caption{The data (left), model (left-middle), residuals (right-middle) and the reconstructed source (right) for the high resolution Stokes I data from \datares{} (first row), and the polarization data from from \datapol{}, in Stokes Q (second row) and U (third row). The critical lines and caustics are shown as white curves. The data and model are in units of m\unitjybeam{} (first row), $\mu$\unitjybeam{} (second and third row) and the residuals are in units of $\sigma$. For the reconstructed source, the units are normalized to the peak surface brightness.}
    \label{fig:polmodels}
\end{figure*}

We jointly infer $\source{}$ and $\lensparam{}$ using a hierarchical Bayesian framework. At the first level of inference we find the maximum a posteriori source, $\sourcemap{}$, by maximizing the probability
\begin{equation}\label{eq:infersource}
    P(\source{} \,|\, \data{}, \lensparam{}, \lambdas{}, \regul{})
    = \frac
    {
    P(\data{} \,|\, \source{}, \lensparam{}) \,
    P(\source{} \,|\, \lambdas{}, \regul{})
    }
    {
    P(\data{} \,|\, \lambdas{} , \lensparam{}, \regul{})
    }\,,
\end{equation}
where we use a regularizing prior $P(\source{} \,|\, \lambdas{}, \regul{})$ of strength $\lambdas{}$ and order $\regul{}$ to penalize models where the source has strong surface brightness gradients indicative of poorly-fitting lens parameters. We note that in the case of polarized data the regularization strength $\lambdas{}$ is a vector of length $N_{\rm corr}$, but a scalar quantity when modelling only Stokes I. We maximize the posterior in Equation (\ref{eq:infersource}), assuming a quadratic form of $P(\source{} \,|\, \lambdas{}, \regul{})$ with zero mean, by solving the following system of linear equations,
\begin{equation}
    \bmath{\mathsf{A}} \,\sourcemap{} = (\dirty{}\lensop{})^T\invcov{}\data{}\,,
\end{equation}
where
\begin{equation}
    \bmath{\mathsf{A}} = \left[(\dirty{}\lensop{})^T \invcov{} \dirty{}\lensop{} + \lambdas{} \regul{}_{\source{}}^T\regul{}^{}_{\source{}}\right]\,,
\end{equation}
using a preconditioned conjugate gradient solver \citep{powell_2021}.
At the second level of inference we obtain the maximum a posteriori lens parameters $\lensparam{}$ and the regularization strengths $\lambdas{}$ by maximizing
\begin{equation}\label{eq:secondlevel}
\begin{split}
    2P(\data{} \,|\, \lensparam{}, \lambdas{}, \regul{})
    = &-\chi^2 - \lambdas{}\sourcemapT{}\regul{}^T_{\source{}}\regul{}^{}_{\source{}}\sourcemap{} - \log\det \bmath{\mathsf{A}}  \\
    &+ \log\det\left(\lambdas{}\regul{}^T_{\source{}}\regul{}^{}_{\source{}}\right) + \log\det \left(2\pi \invcov \right)\,,
\end{split}
\end{equation}
where
\begin{equation}
    \chi^2 = \left( \dirty{} \lensop{} \sourcemap{} - \data{} \right)^T \invcov{} \left( \dirty{} \lensop{} \sourcemap{} - \data{} \right)\,.
\end{equation}
At the final level of inference we use \multinest{} \citep{feroz_2009} in order to obtain the full posterior distribution of all free parameters.

\subsection{Lens Model}\label{sec:model}
The \HST{} data show that the main lensing galaxy (G1) is accompanied by a fainter satellite galaxy (G2) close to the Einstein ring, which is likely to be at the same redshift \citep{geach_2015}. We model galaxy G1 with a power law elliptical mass density (PEMD) profile \citep{keeton_2001}, which has a normalized projected mass density
\begin{equation}
         \kappa(x, y) = \frac{\kappa_0 \, (2 - \frac{\gamma}{2}) \, q^{\gamma - \frac{3}{2}}}
         {2\left[q^2 \, x^2 + y^2 \right]^{\frac{\gamma -1}{2}}}\,,
\end{equation}
where $\kappa_0$ is the mass normalization factor, $q$ is the axis ratio and $\gamma$ is the density slope.

We include an external shear component in the model for G1, parameterized with an amplitude $\Gamma$ and direction $\theta_{\Gamma}$. For the companion galaxy we use an isothermal sphere model with $\kappa_0=0.1079$ \citep{liu_2022}. We fix the positional offset between G1 and G2 to $\Delta x = 0.45$ and $\Delta y = 2.31$~arcsec, as measured from their near-infrared positions in the \HST{} image.

To account for the possibility of additional angular complexity in the lens, we extend the lens model of G1 to include multipoles, allowing the mass distribution to deviate smoothly from a strictly elliptical form. Following the notation from \citet{oriordan_2024}, the additional convergence terms due to multipoles are
\begin{equation}
        \kappa_{\rm m}(r, \phi) = r^{(1-\gamma)}\left[a_{\rm m}\sin (m\phi) + b_{\rm m} \cos (m\phi)\right]\,,
\end{equation}
where $m$ is the order of the multipole, $\gamma$ is the mass density slope of the main lens, and $r$ and $\phi$ are polar coordinates on the image plane.

We obtain a lens model using the high resolution Stokes I data from \datares{}. We then apply the model for \datares{} to the Stokes IQUV in \datapol{} by keeping the lens model fixed, but allowing the regularization strength and model position to change. We compute the polarization angle and linearly polarized intensity from the Q and U maps, the former of which is rotated by 90 deg to infer the magnetic field position angle.

\section{The reconstructed source}
    \begin{table*}
    	\centering
    	\caption{The lens parameters for G1 as inferred from \datares{}.}
    	\label{tab:lensparams}
    	\begin{tabular}{lllllllllllll}
    		\hline
    		 & $\kappa_0$ & $\posang{}$ ($^{\circ}$) & $q$ & $x$ ($\arcsec$) & $y$ ($\arcsec$) & $\gamma$ & $\Gamma$ & $\theta_\Gamma$ ($^{\circ}$) & $a_3$ & $b_3$ & $a_4$ & $b_4$  \\
    		\hline
                Mean & $2.419$ & $85.07$ & $0.9323$ & $-0.3794$ & $0.2362$ & $1.934$ & $0.05950$ & $75.25$ & $-0.0107$ & $0.0998$ & $0.0023$ & $0.0054$  \\
                $\sigma$ & $0.019$ & $1.65$ & $0.0058$ & $0.0051$ & $0.0028$ & $0.015$ & $0.0019$ & $0.42$ & $0.0028$ & $0.0015$ & $0.0025$ & $0.0027$  \\
    		\hline
        \end{tabular}
    \end{table*}

    In Fig.~\ref{fig:polmodels} and Table~\ref{tab:lensparams}, we present the results of modelling the two data sets that have been used for our analysis. Fig.~\ref{fig:polmodels} shows the lens modelling results for the total intensity dust continuum of data set \datares{} and for the Stokes Q, and U emission for data set \datapol{}. While \pronto{} fits the data directly in visibility space, we present the Fourier transform of the data and the residuals for visualization purposes, where the residuals are computed according to \begin{equation}
        \rm residual\,\,=\,\, \frac{data \,\,-\,\,model}{\sigma_{rms}}.
    \end{equation}We find that both data set \datares{} and Stokes Q and U from \datapol{} can be well-modelled to the noise level, although the residuals for \datares{} show that the model is slightly overfitting. Table \ref{tab:lensparams} shows the inferred lens parameters for the main lensing galaxy, G1. A further discussion of the mass properties of the lens and its environment will be presented in a forthcoming paper.

\begin{figure}
        \includegraphics[width=1\columnwidth]{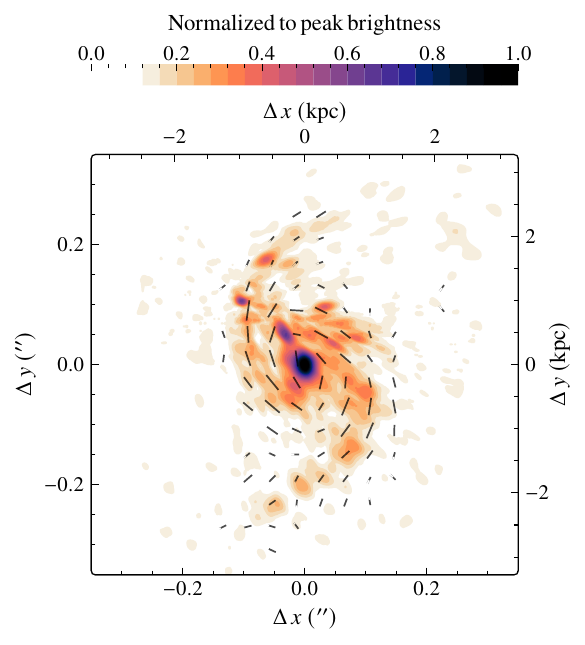}
        \caption{A colour map showing the surface brightness distribution of the reconstructed heated dust from \datares{}. The orientation of the inferred magnetic field in the source plane is indicated by the black quivers, with lengths proportional to the linearly polarized intensity. The coordinate system on the top and right axes indicate a physical scale in kpc at $z=2.553$, and the colour-bar shows the surface brightness normalized to the peak intensity.}
        \label{fig:bfield}
    \end{figure}

    In Fig.~\ref{fig:bfield}, we present the reconstructed source plane model of the heated dust emission (from data set D$_1$) with the magnetic field positional angle quivers (from data set D$_2$) overlayed. This model has structures resolved on around 20 to 100 pc in the heated dust continuum, although the effective angular resolution of data set D$_1$ varies across the source plane due to the differential magnification. The reconstruction reveals that the morphology of the heated dust resembles a grand-design spiral with two arms and a bright central core, possibly with a bar. The spiral arms have an extent on the order of 4 kpc in the north--south direction and about 2 kpc in the east--west direction, and they contain several bright, compact regions. The brightest of these is found roughly 500 pc to the north-east of the galaxy core, close to the caustic line. These regions of bright emission are likely sites of ongoing and intense star-formation within the host galaxy of \src{}.

    The quivers indicating the magnetic field orientation, which have an angular uncertainty of $\leq 5 \deg$, clearly follow the spiral structure of the arms, which is also where the linearly polarized intensity is strongest. The spacing between quivers (40 mas) is larger than the effective angular resolution in the source plane at all regions of interest and hence the individual quivers can be assumed to be uncorrelated. Other studies of DSFGs at mm-wavelengths with ALMA have found that emission from the dust continuum is often clumpy and compact, concentrated in the central 1 to 2 kpc, and generally does not trace the full extent of the stellar distribution \citep[e.g.][]{rybak_2015, fujimoto_2017,gomez-guijarro_2022,stacey_2024}. We note that a recent analysis of lensed DSFGs have found evidence for extended structures in the reconstructed sources, with morphologies that are consistent with spiral arms \citep{stacey_2024,stacey_2025}. Finally, we note that a second source at the same redshift as \src{} was identified by \citet{liu_2022}, who suggested that the two may be undergoing a merger. We find no source plane emission in either of the two data sets at the location that \citet{liu_2022} predicted for the second source, but it is clear from the full Einstein ring seen in Fig.~\ref{fig:hubblealma} that faint dust continuum emission extends well beyond what we can recover in our current reconstruction. In a future paper, we will combine all of the available data for \src{} to build a complete picture of the multi-wavelength properties of this object.

\section{Discussion and Conclusion}

    It is widely accepted that magnetic fields in galaxies originate from primordial seed fields in the early Universe \citep{kulsrud_2008}, and have since been amplified by turbulent dynamos. In contrast to fluctuation dynamos that occur on small scales \citep{brandenburg_2005}, a mean-field dynamo not only amplifies the field, but also organizes it into a large-scale ordered structure; a process that is driven by differential rotation of the galactic disk \citep{ruzmaikin_1988}. While fluctuation dynamos alone can quickly amplify weak seed fields to strengths close to equipartition even before a galaxy first forms a disk  \citep{schekochihin_2004,brandenburg_2005,schober_2012,pakmor_2014,pakmor_2024}, the saturated field it generates will be completely chaotic. Once the galaxy forms a differentially rotating disk the magnetic field will be dragged along with the rotation and slowly ordered on timescales of several orbits. However, it will keep many field reversals. In contrast, the large-scale ordering of the field by the mean-field dynamo is estimated to take several billion years, but will lead to an ordered, large-scale field \citep{beck_2013, rodrigues_2019}. Hence, any observations of ordered fields in galaxies at $z\gtrsim 3$ can provide important clues on how magnetic fields evolve over cosmic time, on their amplification mechanisms and time-scales \citep{beck_2013}, and potentially also about the age of the gas disk.

    In the case of \src{}, we have clearly detected evidence of a highly ordered magnetic field in a grand-design spiral when the Universe was at an age of just 2.6 Gyr. We see that the inferred magnetic field has a kpc-scale ordered structure and is aligned with the spiral arms, as traced by the heated dust emission. While this result differs from that of \citet{geach_2023}, who have assumed a uniform polarization angle across the source, we find no indication that the two results are inconsistent. This represents the second resolved detection of an ordered magnetic field in the early Universe, but unlike in the case of SPT 0346--52 \citep{chen_2024}, the structure is well-resolved and the lensing configuration provides a complete picture of the source plane morphology. For SPT 0346--52, the ordered magnetic field is patchy, is offset from the heated dust and is more closely associated with a component of the [C~{\sc{ii}}] emission. This suggested that the magnetic field is complex and possibly disturbed by a merging event. However, although SPT 0346--52 lies at the cosmologically interesting epoch of redshift 5.6, the lensing configuration is dominated by two compact lensed images, which provide only limited information to constrain the lens mass model, and the magnetic field information is mainly restricted to a highly magnified part of the source, which biases the interpretation. Neither of these issues are at play in the case of \src{}, where a morphology that is strikingly consistent with observations of magnetic fields in nearby spirals is seen \citep[e.g.][]{han_1999, fletcher_2011}.

     The results presented here raise a question about the timescales involved in the ordering of galactic magnetic fields. With differential rotation of the disk being seen as one of the main drivers of winding-up, as well as the mean-field dynamo \citep{ruzmaikin_1988}, one may reasonably expect that organizing the field requires at least a certain amount of rotations of the disk. Since winding up of the magnetic field will retain many field reversals, future constraints on those might be able to tell us if a large-scale dynamo has already contributed significantly to shaping the magnetic field of the galaxy, and potentially constrain the time since the formation of the gas disk.

    The recent detections of polarized emission in the early Universe by \citet{geach_2023} and \citet{chen_2024} present an exciting avenue for studying the magnetic field distribution of DSFGs with ALMA. However, this technique will only test galaxy formation models when high resolution imaging ($<200$~mas) is combined with sophisticated gravitational lens modelling methods, like those within \pronto{} \citep{ndiritu_2024}. From total intensity reconstructions of lensed DSFGs, there is growing evidence for complex morphologies on 20 to 100 pc-scales that resemble spiral-arm structures \citep{stacey_2024,stacey_2025}. Therefore, observing other targets should be feasible, which is clearly important for determining whether the highly ordered magnetic field observed in the case of \src{} is the norm or the exception.

\section*{Acknowledgements}
S.~V. thanks the Max Planck Society for support through a Max Planck Lise Meitner Group. Part of this research was carried out on the High-Performance Computing resources of the FREYA cluster at the Max Planck Computing and Data Facility (MPCDF) in Garching, operated by the Max Planck Society (MPG). This work is based on the research supported in part by the National Research Foundation of South Africa (Grant Number: 128943). This paper makes use of the following ALMA data: ADS/JAO.ALMA\#2023.1.01585.S and ADS/JAO.ALMA\#2021.1.00458.S. ALMA is a partnership of ESO (representing its member states), NSF (USA) and NINS (Japan), together with NRC (Canada), MOST and ASIAA (Taiwan), and KASI (Republic of Korea), in cooperation with the Republic of Chile. The Joint ALMA Observatory is operated by ESO, AUI/NRAO and NAOJ.

\section*{Data Availability}

The data used here is publicly available from the ALMA archive.



\bibliographystyle{mnras}
\bibliography{references} 





\bsp	
\label{lastpage}
\end{document}